\documentclass[aps,prd,showpacs,superscriptaddress,onecolumn,raggedfooter,raggedbottom]{revtex4-1}   

\usepackage{amssymb}
\usepackage{amsmath}
\usepackage{amssymb}
\usepackage{graphicx}
\usepackage{subfigure}
\usepackage{color}
\usepackage{mathrsfs} 
\usepackage{float}

\usepackage{soul}

\usepackage{hyperref}

\begin{document}

\title{Kink-Antikink Interaction Forces and Bound States in a Biharmonic $\phi^4$ Model}

\author{Robert J.\ Decker}
\affiliation{Mathematics Department, University of Hartford, 200 Bloomfield Ave., West Hartford, CT 06117, USA}

\author{A.\ Demirkaya}
\affiliation{Mathematics Department, University of Hartford, 200 Bloomfield Ave., West Hartford, CT 06117, USA}

\author{N.~.S.\ Manton}
\affiliation{Department of Applied Mathematics and Theoretical Physics, University of Cambridge, Wilberforce Road, Cambridge CB3 0WA, U.K.}

\author{P.~G.\ Kevrekidis}
\affiliation{Department of Mathematics and Statistics, University of Massachusetts,
Amherst, MA 01003-4515, USA}

\affiliation{Mathematical Institute, University of Oxford, Oxford, UK}

\begin{abstract}
We consider the interaction of solitons
in a biharmonic, beam model analogue of the
well-studied $\phi^4$ Klein-Gordon theory. Specifically, 
we calculate the force between a well separated kink and antikink.
Knowing their accelerations as a function
of separation, we can determine their motion 
using a simple ODE. There is good agreement between
this asymptotic analysis and numerical computation.
Importantly, we find the force has an
exponentially-decaying oscillatory behaviour
(unlike the monotonically attractive interaction in the Klein-Gordon
case). Corresponding to the zeros of the force, we predict the
existence of an infinite set of field theory equilibria, i.e., kink-antikink bound
states. We confirm the first few of these at the PDE level, and verify 
their anticipated stability or instability.
We also explore the implications of this interaction force in the
collision
between a  kink and an oppositely moving antikink.

\end{abstract}

\maketitle

\section{Introduction}

The symmetry-breaking $\phi^4$ potential has a time-honoured history in the context of nonlinear
partial differential equations (PDE), especially of the Klein-Gordon type~\cite{belova,campbell2}.
In nonlinear Klein-Gordon theory, the interaction and collisions of kinks and antikinks remains a
somewhat elusive topic~\cite{roy}, and research into this is (still) ongoing~\cite{weigel2,clisthenis}. 
This theory combines a Laplacian with a $\phi^4$ potential, and it is well-known that kinks and 
antikinks attract~\cite{Manton_nuclear}. Yet the interplay of translational, internal
and extended (phonon) modes at relatively 
high speeds~\cite{Sugiyama,Campbell,Ann,goodman,goodman2,weigel}, leading
to fractal, so-called multi-bounce collision windows, still eludes a self-consistent, low-dimensional 
effective particle description. The reader is referred to~\cite{cuevas} for a summary of
recent developments on the subject.

The standard $\phi^4$ Klein-Gordon theory yields the field equation
\begin{equation}\label{phi4}
u_{tt}=u_{xx}-V'(u) \,, 
\end{equation}
where $V(u)=\frac{1}{2}(u^2-1)^2$. In this paper, we further explore a variant,
referred to as the nonlinear beam model~\cite{beam_demirkaya,beam1},  Here, the 
harmonic spatial derivative term is replaced by a biharmonic term, and the field equation is
\begin{equation}\label{beam}
u_{tt}=-u_{xxxx}-V'(u) \,,
\end{equation}
with $V(u)$ as before.

Similar variant models have been recently considered
by a number of authors~\cite{levandosky,champneys,CM,karageorgis}.
They have potential applications to the propagation of travelling waves in suspension 
bridges; there, the models often involve piecewise constant or exponential nonlinearities.
Part of our interest stems from a recent development in the realm
of nonlinear Schr{\"o}dinger (NLS) equations, of which the real-field
equation considered here is a simplification. In particular, in the
context of nonlinear optics, the possibility of the so-called pure quartic
solitons has {\it experimentally} showcased the potential of quartic dispersion
combined with cubic nonlinearities, similar to what we study here~\cite{pqs}.
Moreover, a very recent extension has considered combining
harmonic and biharmonic terms~\cite{pqs2}, and it is interesting to note that 
linearized  models of this mixed type occur in the context of stiff strings 
and piano tuning~\cite{gracia}. Finally, the existence
and stability of standing waves in certain NLS models~\cite{atanas} 
may be connected with the real field phenomena found here.

In earlier work~\cite{beam_demirkaya,beam1}, some of the present authors 
explored the existence, asymptotic tail properties, and stability of both static 
and travelling single kinks, and complemented this with a numerical investigation of 
kink-antikink collisions as a function of the incoming speeds. We found no 
multi-bounce windows or accompanying fractal structure. We did, however, 
find an intriguing oscillatory behaviour in the velocity-out (i.e., outgoing
velocity) versus velocity-in (incoming velocity)
graph at the boundaries of the bound state interval of velocity-in values.

Here, we extend our study of the dynamics of kinks and antikinks in this beam model -- a
biharmonic nonlinear field theory. Specifically, we first calculate the asymptotic force
and associated interaction potential between a kink and antikink, using
the method of~\cite{Manton_nuclear}. Using our knowledge that the single kink 
(and antikink) tails are spatially oscillatory and exponentially decaying~\cite{beam1},
we derive an explicit formula for the force. ln contrast to the 
harmonic case, we find that as a function of separation the sign of the force 
alternates between a sequence of zeros. The force is 
{\it not} universally attractive as in harmonic field theories, 
but rather alternates between attractive and repulsive.
The consequence is the existence of a sequence of equilibria, i.e.,
bound states of a kink and antikink. A topological constraint forces
these equilibria to alternate between local maxima and minima in the potential  energy
landscape, i.e., between saddles and centers of the associated
dynamical system. There is a self-similar pattern of
progressively (exponentially) smaller basins between adjacent saddles
where the kink and antikink can be trapped in an oscillatory motion. 
We then confirm these predictions by full eigenvalue computations around
the equilibria, and also by solving the dynamical PDE for the field. The
features we have discovered are novel, to the best of our knowledge,
and it is interesting to explore if they persist
in settings involving mixed harmonic and biharmonic terms (and, of
course, beyond Klein-Gordon models).

In Sec.~II, we present the basic mathematical features of the beam model --
the Lagrangian and Hamiltonian, and the conservation laws  of energy and 
momentum -- and explain how to adapt
the ideas of~\cite{Manton_nuclear} to the present biharmonic setting.
We then derive a formula for the acceleration of a well-separated kink and antikink, as a function 
of their separation. In Sec.~III, we present results of a systematic numerical investigation
of the kink-antikink solutions of the field theory PDE, and compare the asymptotic, analytical predictions. We also 
examine the implications for kink-antikink collisions. Finally, in Sec.~IV,
we summarize our findings and outline some future challenges.

\section{Theory of the Kink-Antikink Interaction}

 For our nonlinear beam model, the Lagrangian density  is 
\begin{equation}\label{lagr1}
\mathcal{L}(u;t) = \mathcal{T}(u;t) - \mathcal{V}(u;t) =
\frac{1}{2}u_t^2 - \left(\frac{1}{2}u_{xx}^2 + V(u)\right) \,,
\end{equation}
and the Lagrangian is  
\begin{equation}
L= \int_{-\infty}^\infty {\cal L} dx= \int_{-\infty}^\infty  \left(\frac{1}{2}u_t^2 - \frac{1}{2}u_{xx}^2 -
V(u)\right) \, dx \,,
\end{equation}
leading to the field equation Eq.~(\ref{beam}). Naturally, the corresponding Hamiltonian is
\begin{equation}\label{lagr}
\mathcal{H}(u;t) = \int_{-\infty}^\infty \left(\mathcal{T}(u;t) + \mathcal{V}(u;t)\right) dx
= \int_{-\infty}^\infty  \left(\frac{1}{2}u_t^2 + \frac{1}{2}u_{xx}^2 +
V(u)\right) \, dx \,.
\end{equation}

The momentum on the interval $[x_1,x_2]$ is given by the standard expression
\begin{equation} \label{momentum}
P=-\int_{x_1}^{x_2} u_t u_x \, dx \,.
\end{equation}
When $x_1$ and $x_2$ tend to $-\infty$ and $\infty$, $P$ is the total momentum, and 
using the field equation, one can show that this is conserved. Here, we
will instead use Eq.~(\ref{momentum}) in a more limited spatial range, in the spirit of the calculation
of~\cite{Manton_nuclear}, in order to calculate the force that a kink exerts on an antikink.

Differentiating $P$ with respect to time $t$, and using Eq.~(\ref{beam}), we find that
\begin{align}
\frac{dP}{dt}=-&\int_{x_1}^{x_2} \left(u_{tt}u_x+u_t u_{xt}\right) \, dx 
\nonumber \\
=&\int_{x_1}^{x_2} 
\left(u_{xxxx}u_x+V'(u)u_x-\frac{1}{2}(u_t^2)_x\right) \, dx \nonumber \\
=&\left[u_xu_{xxx}-\frac{1}{2}u_{xx}^2+V(u)-\frac{1}{2}u_t^2\right]_{x_1}^{x_2}  \,,
\label{eq:momentum}
\end{align}
where the quantity in square brackets is the component $T_{xx}$ of the 
energy-momentum tensor~\cite{manton_sutcliffe}. The last expression
can be interpreted as the 
force $F$ acting on the part of the field between $x_1$ and $x_2$.
For a field configuration $u(x,t) = \varphi(x)$ that is static or almost 
so, we can ignore the term involving $u_t^2$, and the force becomes
\begin{equation}
F = \left[\varphi_x \varphi_{xxx} - \frac{1}{2}\varphi_{xx}^2 + 
V(\varphi) \right]_{x_1}^{x_2}.
\end{equation}
The quantity in square brackets is now the first integral of the static 
field equation
$\varphi_{xxxx} + V'(\varphi) = 0$, so it is a constant, independent of 
$x$, if $\varphi(x)$ satisfies this equation. Therefore there is no 
force acting on any part of an exact static solution,
consistent with the momentum $P$ of such a state being zero and
remaining so. However, we are interested in the non-zero force for a kink-antikink 
configuration $\varphi(x)$ that is only static instantaneously.

So, consider a concrete field configuration $\varphi(x)$ that is a superposition 
of a kink solution centered
at $-X$ and an antikink centered at $X$, where $X$ is large and positive 
so the antikink-kink separation
$2 X$ is large. The fields of the individual kink and antikink are 
$\varphi_K(x+X)$ and
$\varphi_{AK}(x-X) = -\varphi_K(x-X)$, where $\varphi_K(x)$ denotes the 
kink centered at the origin. Their superposition is
\begin{equation}
\varphi(x) = \varphi_K(x+X) + \varphi_{AK}(x-X) - 1 \,.
\end{equation}
The shift by $-1$ is required to satisfy the boundary conditions 
$\varphi(x) \to -1$ as $x \to \pm\infty$.

In the region between the kink and antikink, near $x=0$, $\varphi(x)$ is 
a superposition of the kink and antikink tail fields. Let us write 
$\varphi_K(x) = 1 - \eta_K(x)$. For large positive $x$, the kink tail 
$\eta_K(x)$ is spatially oscillatory and exponentially small. Its precise form is $\eta_K(x) = b 
e^{-x} \cos(x-d)$, where the parameters have been 
determined numerically in \cite{beam1} to be $b \approx 0.9650$ and $d 
\approx 0.4086$. Then, in the region between the kink and antikink we 
can write $\varphi(x) = 1 - \eta(x)$ where
\begin{equation}
\eta(x) = \eta_K(x+X) + \eta_{AK}(x-X) \,.
\end{equation}
$\eta_{AK}$, the tail of the antikink (to its left) is the reflection of 
$\eta_K$, the tail of the kink (to its right).

To find the force on the antikink, due to the kink, we need to evaluate 
for the field configuration $\varphi(x)$ the expression $F$ above, 
setting $x_1 = 0$ and $x_2 \rightarrow \infty$. The contribution from
$x_2$ vanishes, as the field derivatives all vanish there, and so does $V$ 
because $\varphi$ satisfies the boundary conditions. At $x_1 = 
0$, $\varphi$ differs from 1 by the sum of the exponentially small 
tails, so we can replace $V(\varphi)$ by $V(1 - \eta) \simeq 2\eta^2$. 
The derivatives of $\varphi$ are minus the derivatives of $\eta$, so the 
force simplifies to the quadratic expression
\begin{equation}
F = -\eta_x \eta_{xxx} + \frac{1}{2}\eta_{xx}^2 - 2\eta^2 \,.
\end{equation}

The right hand side of $F$ is now the first integral of the linearized 
static field equation, $\eta_{xxxx} + 4\eta = 0$. The kink tail $\eta_K$ 
satisfies this equation, and also decays exponentially as $x$ increases, 
so for $\eta_K$ by itself the force is zero; similarly so for the antikink 
tail $\eta_{AK}$, which decays exponentially as $x$ decreases. These 
self-forces can also be shown to be zero by direct calculation. For 
$\eta = \eta_K + \eta_{AK}$, it is therefore only the cross terms (the 
interaction terms) that give a non-zero force, so
\begin{equation}
F = -(\eta_K)_x (\eta_{AK})_{xxx} - (\eta_K)_{xxx} (\eta_{AK})_x + 
(\eta_K)_{xx} (\eta_{AK})_{xx}
- 4(\eta_K)(\eta_{AK}) \,.
\end{equation}
Note that since $\eta(x)$, the sum of the tails, also satisfies the linearized static field equation, 
this force is independent of where it is evaluated in the region between 
the kink and antikink. For convenience, we are evaluating it at $x=0$.

The tail of the kink centered at $-X$ is $\eta_K(x+X) = b e^{-(x+X)} \cos(x+X-d)$, 
and its derivatives are
\begin{align}
(\eta_K)_x =& -b e^{-(x+X)} (\cos(x+X-d) + \sin(x+X-d)) \nonumber \\
(\eta_K)_{xx} =& 2b e^{-(x+X)} \sin(x+X-d) \nonumber \\
(\eta_K)_{xxx} =& 2b e^{-(x+X)} (\cos(x+X-d) - \sin(x+X-d)) \,.
\end{align}
The tail of the antikink centred at $X$ is $\eta_{AK}(x-X) = b e^{x-X} \cos(x-X+d)$, 
and its derivatives are similar. Combining the results for $\eta_K$ and 
$\eta_{AK}$, and using a trigonometric addition formula,
we find, finally, that the force that the kink exerts on the antikink is
\begin{equation}
F = -8 b^2 e^{-2 X} \cos(2 X - 2d) \,.
\end{equation}
The kink at $-X$ experiences the opposite force.

The inertial mass of a single kink or antikink can be found from its momentum 
$P$. Suppose a kink is centered at the moving point $X(t)$ and that $\dot{X}$ is 
small, so the kink profile is approximately that of a static kink. 
Then $u(x,t) = \varphi_K(x + X(t))$, and from Eq. (\ref{momentum}) we see that the 
kink momentum is $M {\dot{X}}$, where
\begin{equation}
M = \int_{-\infty}^{\infty} (\varphi_K)_x^2 \, dx \,.
\end{equation}
A similar calculation of the kinetic energy of a moving kink gives 
$\mathcal{T} = \frac{1}{2} M {\dot{X}}^2$. Numerically, it has been 
determined that $M \approx 1.1852$. Note that $M$ is not the static 
energy of the kink; this is consistent in a theory without Lorentz 
invariance (contrary, e.g., with the situation in the nonlinear
Klein-Gordon models such as Eq.~(\ref{phi4})).

The equation of motion for the antikink is therefore
\begin{equation}
M {\ddot{X}} = -8 b^2 e^{-2 X} \cos(2 X - 2d) \,.
\end{equation}
The separation $s = 2 X$ obeys the equation $\frac{1}{2} M {\ddot s} = -8 
b^2 e^{-s} \cos(s - 2d)$; as usual for two bodies of equal mass $M$, 
this involves the reduced mass $\frac{1}{2} M$. Using the parameter 
values $M, b$ and $d$ given above, we find the acceleration of the 
antikink is
\begin{equation}
\label{prediction}
{\ddot{X}} = - 6.286 e^{-2 X} \cos(2 X - 0.8172) \,.
\end{equation}
This asymptotic analytical result will be compared with the result of a direct
numerical computation in the next section.

\section{Numerical Results}

As in  \cite{beam1} we use Fourier-based spectral methods~\cite{trefethen} to discretize
Eq.~(\ref{beam}) in the spatial direction. Here we use the interval $x\in[-50, 50]$ with an increment of $\Delta x=0.2$. We couple this with Matlab's built-in ODE solver \textit{ode45} to create our PDE simulations, and again \textit{ode45} for the ODE simulations.

\subsection{Kink-Antikink Acceleration and Equilibrium Solutions}\label{steadySec}
In this section, we employ the method developed in \cite{christov} to determine the force
(as measured by the acceleration) between an initially stationary kink and antikink as a
function of $x_{0}$ (half of the separation distance). Similar to
\cite{christov}, we find $\varphi_{\min }(x_{0})$ which minimizes the quantity 
$\||\varphi^{(4)}+V^{\prime }(\varphi)||_2^2$ subject to keeping the positions of the kink and 
antikink (and hence $x_{0}$) constant, using nonlinear least squares ($lsqnonlin$ in Matlab).

For the initial trial input in $lsqnonlin$ we make use of static solutions to
Eq.~(\ref{phi4}). In particular, if $u_{0}(x)$ is a static kink
in the $\phi ^{4}$ Klein-Gordon model (given by $u_{0}(x)=\tanh (x)$) then we use
\begin{equation}\label{steadyInitializer}
u(x)=u_{0}(x+x_{0})+U(x)(-u_{0}(x-x_{0})-u_{0}(x+x_{0}))
\end{equation}
as the initializer (called the split-domain ansatz in \cite{christov}), where $U(x)$ is the Heaviside 
function. This ansatz, which represents a $\phi^4$ kink and antikink separated by a distance 
of $2x_0$, is sufficiently similar to the corresponding configuration of a beam kink and antikink
to converge to the desired result. 

Then we use $\varphi_{\min }(x_{0})$ as the
initial condition (along with zero initial velocity) in Eq.~(\ref{beam}), and 
allow this initial configuration to evolve for a short period of time 
($0.01$ time units). We track the center of the kink $X_{K}(x_0,t)$ (left-side intersection of the 
PDE solution curve $u(x,t)$ with $u=0$) and find that during this time interval the velocity 
$V_{K}(x_0,t)$ of the kink depends nearly linearly on time (we use $X_{AK}(x_0,t)$ and 
$V_{AK}(x_0,t)$ for the position and velocity of the antikink). We then use the slope of the 
velocity versus time graph to measure the initial acceleration of the kink $A_{K}(x_0)$ 
(similarly, $A_{AK}(x_0)$ represents the initial acceleration of the antikink).

\begin{figure}[hbtp]
\begin{center}
\includegraphics[width=0.46\textwidth]{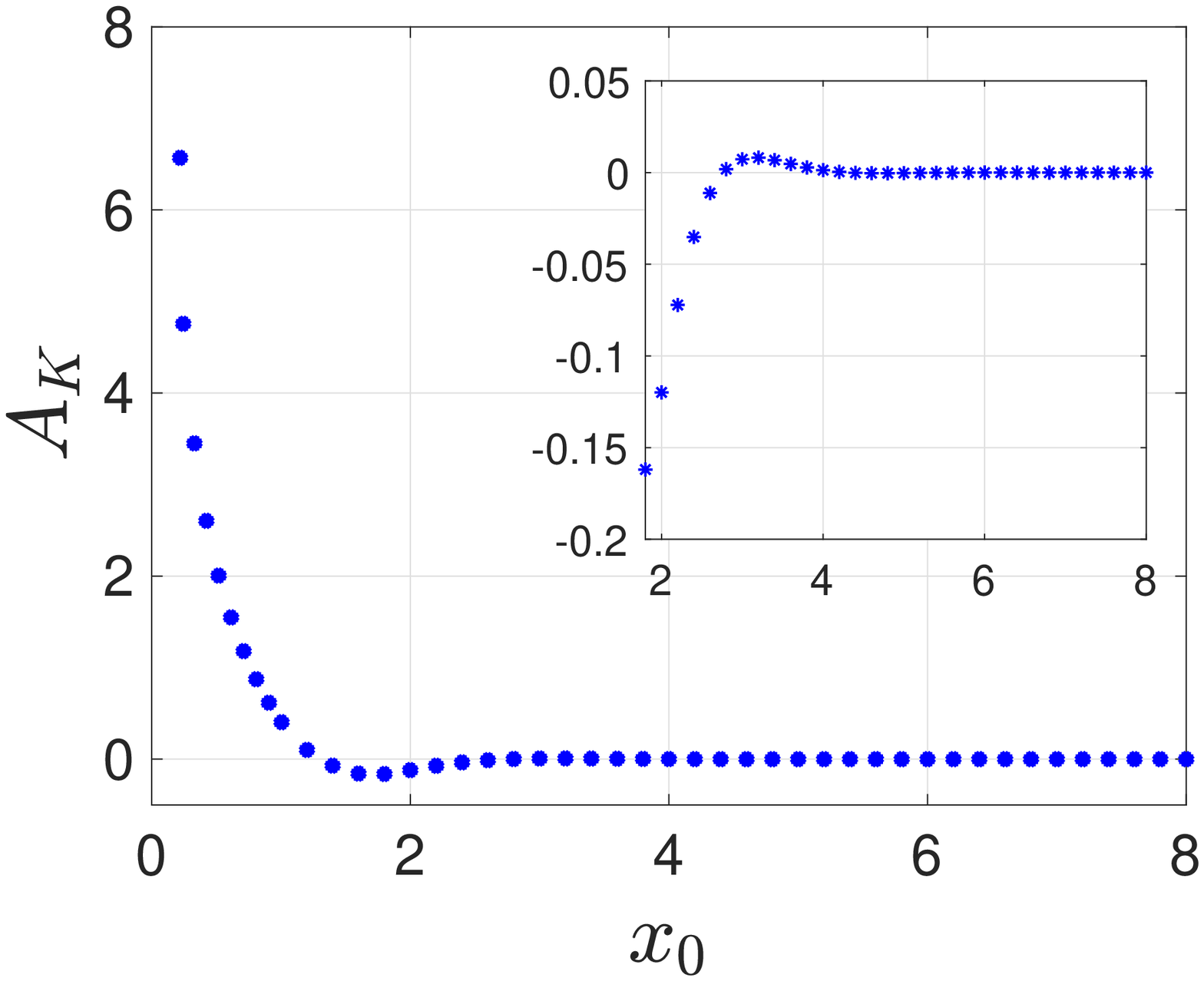}
\includegraphics[width=0.46\textwidth]{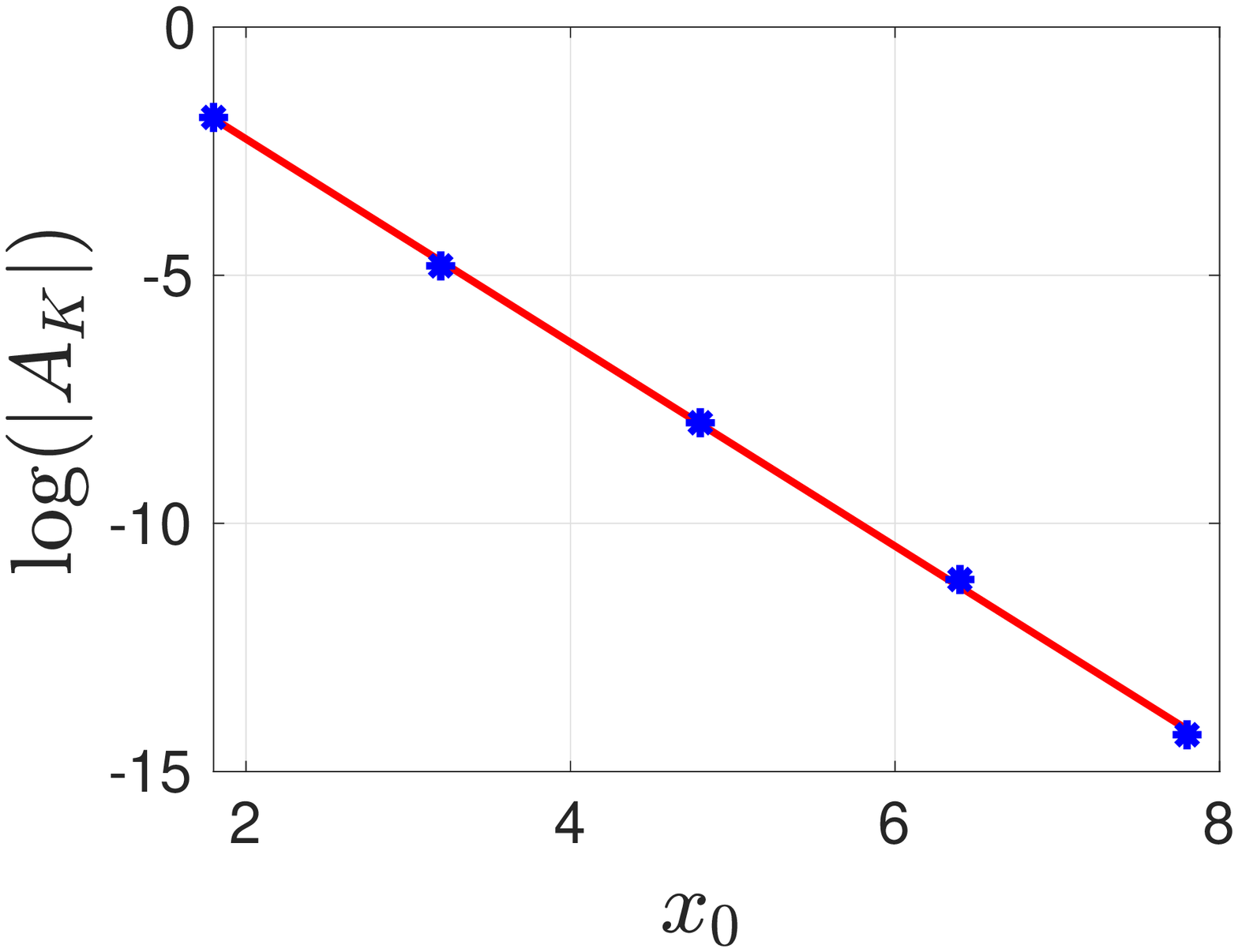}
\includegraphics[width=0.46\textwidth]{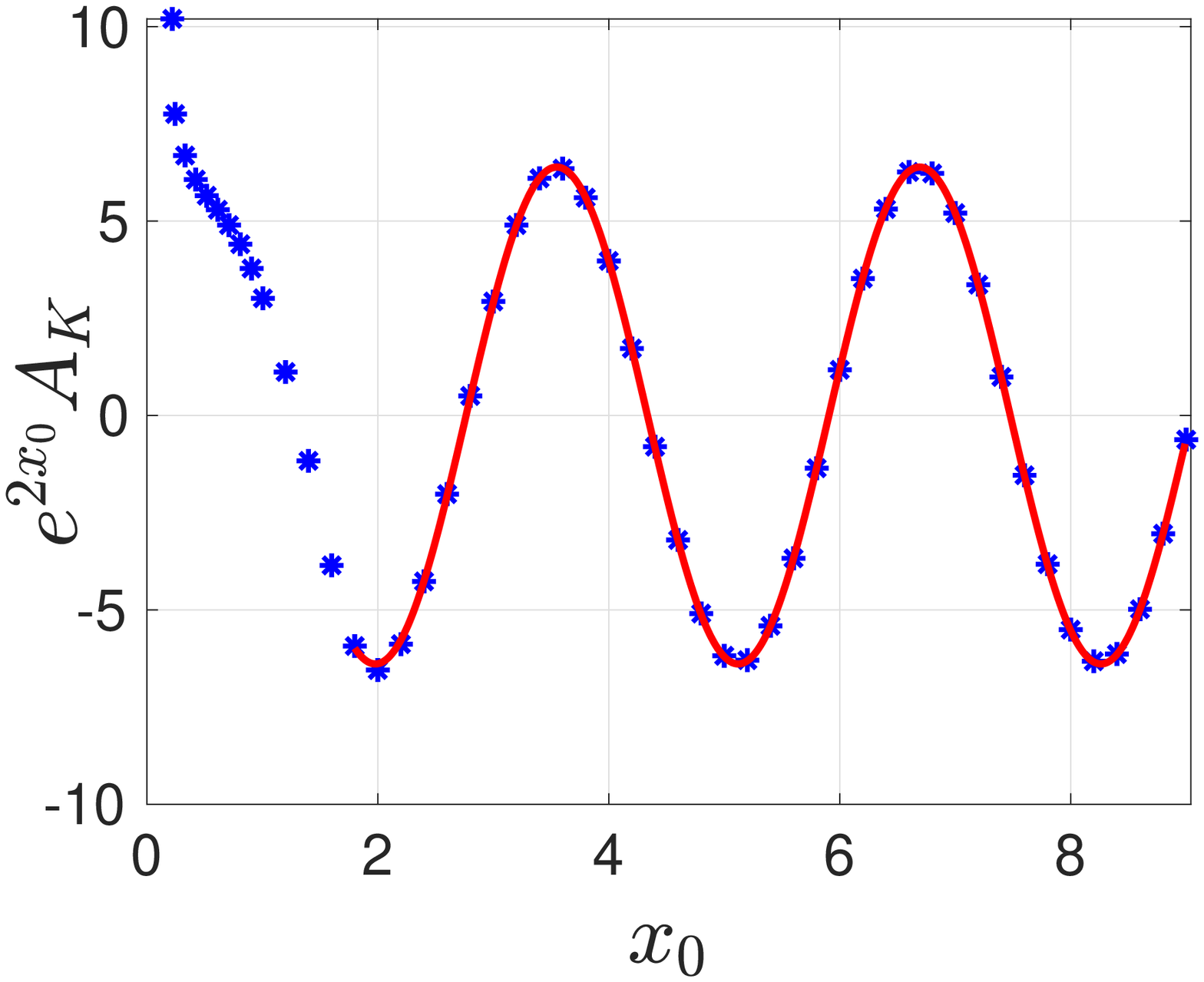}
\includegraphics[width=0.46\textwidth]{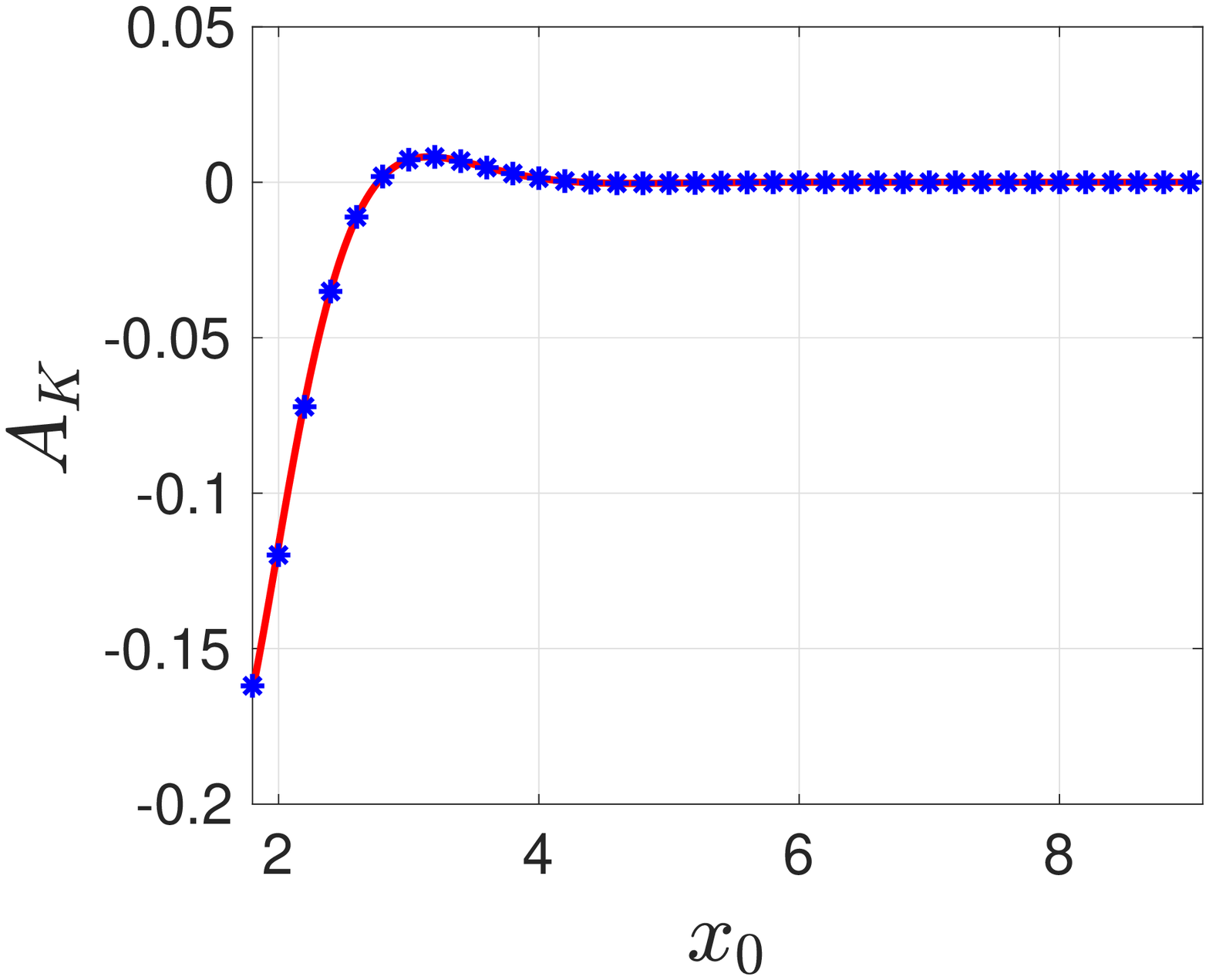}

\end{center}
\caption{Top left panel shows the acceleration of the kink $A_K$ vs half-separation $x_0$. Top right 
panel shows the $x_0$ values of local maxima/minima of acceleration data vs the log of absolute 
value of the acceleration data  (blue stars) and the fitted line $y=-2.051x_0+1.848$ (red solid line). 
Bottom left panel shows  $e^{2x_0}A_K$ vs $x_0$ on the interval $[1.8, 9]$ (blue stars) and the 
fitted curve $y=6.389\cos(2x-0.81590)$ (red solid curve).  Bottom right panel shows  $A_K$ vs 
$x_0$ on the interval $[1.8, 9]$ (blue stars) and the fitted curve $y=6.389e^{-2x_0}\cos(2x_0-0.81590)$ 
(red solid curve).}
\label{kinkForce}
\end{figure}

In Figure \ref{kinkForce}, upper left panel, we show the acceleration of the kink $A_{K}(x_0)$ 
as a function of $x_{0}$ for values of $x_{0}$ in the range $[0.2,9.0]$; in an inset of that figure
we show the same data set, but this time for $x_{0}$ in the range $[1.8,9.0]$. 
Examination of the raw data shows that oscillations continue and that the
acceleration changes sign in regular intervals (for example, the inset shows that
the data becomes positive again in the interval $[2,4]$). Thus the data
appears to have the shape of damped harmonic motion. 
Assuming that the acceleration data may fit a model of the form 
$A_{K}(x_0)=ae^{-bx_0}\cos (cx_0+d)$, we find five data values 
$(x_{0},A_{K}(x_0))$  that represent local maxima or minima of the acceleration, and fit a linear
equation to $(x_{0},\ln (abs(A_{K}(x_0)))$ in order to approximate $b$. The
result is shown in the upper right panel of Figure \ref{kinkForce}. The fit is good with
a slope of approximately $b=-2$, and so we multiply the raw data by $e^{2x_0}$, expecting a 
shifted cosine curve to emerge; we see that this is the case in the plot in the bottom left panel 
of Figure \ref{kinkForce}. This plot indicates that the model is working well for $x_{0}$ values greater
than about $x=1.8$, and so we fit a shifted cosine curve to that part of the data and
lay the fitted curve on top of the data for an excellent fit. For $x_{0}<1.8$, the kink and antikink 
begin to merge and the concept of an acceleration or a force between them loses meaning.

In the bottom right panel of Figure \ref{kinkForce} we show the model 
\begin{equation}
A_K(x_0)=6.389e^{-2x_0}\cos(2x_0-0.81590)
\label{A_k}
\end{equation}
that results from the original acceleration data (with the same data as the inset for the top left
panel). The values of $x_{0}$ where the acceleration is zero should correspond to static equilibria of
Eq.~(\ref{beam}). In between the equilibrium solutions, the kink and antikink
should either approach each other
($A_K(x_0)$ positive, $A_{AK}(x_0)$ negative) or drift apart ($A_K(x_0)$ negative, 
$A_{AK}(x_0)$ positive). This should result in regions of $x_{0}$ values, where steady oscillations 
occur around centers, lying between adjacent saddles in the potential energy landscape. 

In detail, we expect that the motion of the center $X_{AK}(x_0,t)$ of the
antikink will obey the simple ODE (as long as $X_{AK}(x_0,t)\geq 1.8$) 
\begin{equation}\label{akOde}\ddot{X}_{AK}+6.389e^{-2X_{AK}}\cos (2X_{AK}-0.8159)=0 \,. \end{equation}
This is because the antikink has acceleration opposite that of the kink and the position of the
antikink ($X_{AK}(x_0,t)$) is equal to half of the separation between
the kink and antikink (i.e. $X_{AK}(x_0,0)=x_0$). Notice the remarkable agreement of this 
result with the asymptotic prediction of Eq.~(\ref{prediction}). We now further explore the 
validity and implications of this for the nonlinear PDE, Eq.~(\ref{beam}).

\begin{figure}[hbtp]
\begin{center}
\includegraphics[width=0.5\textwidth]{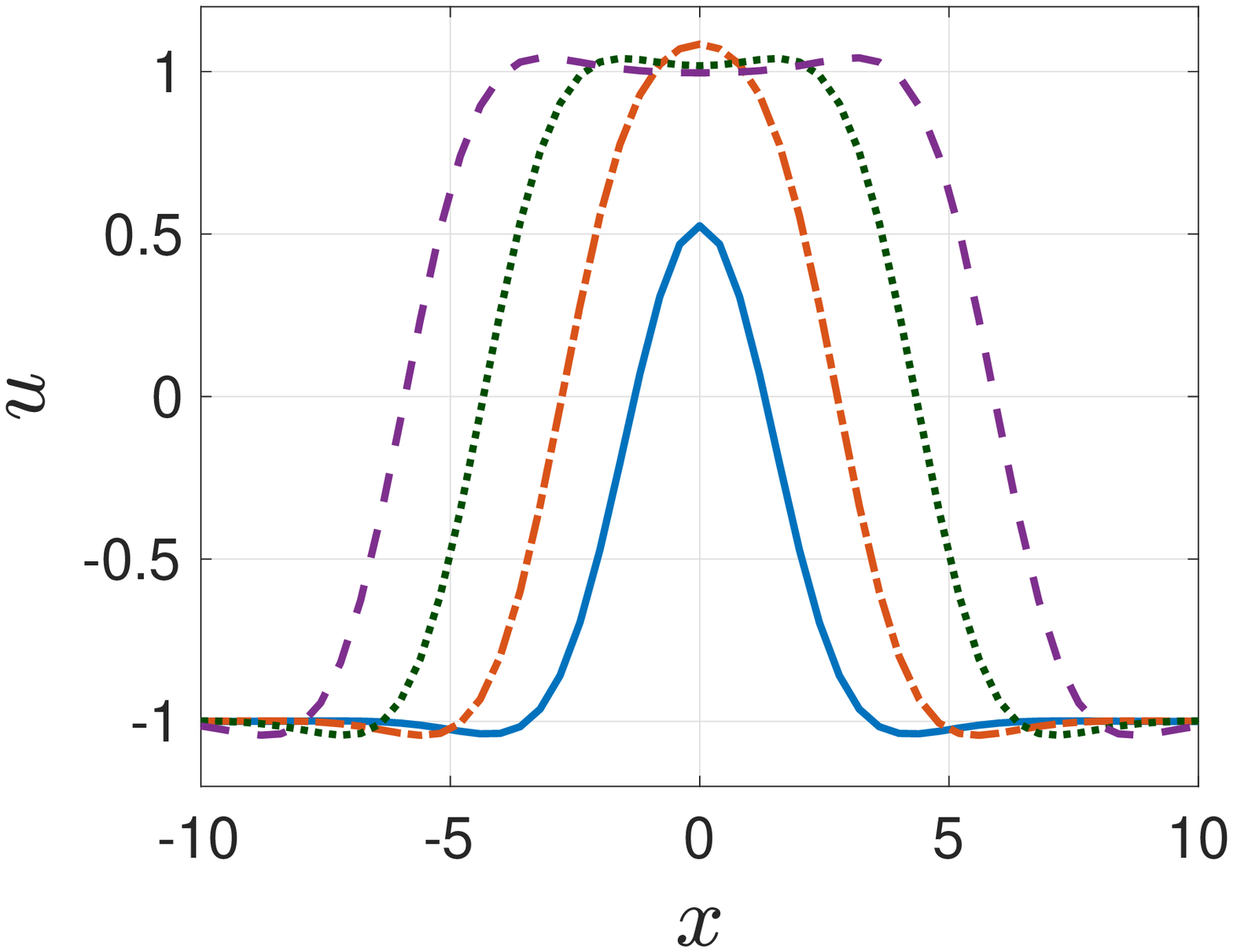}
\end{center}
\caption{Static, equilibrium solutions corresponding to $x_0=1.30$ (blue solid curve), $x_0=2.76$ 
(orange dash-dot curve), $x_0=4.34$ (green dot curve), $x_0=5.91$ (purple dashed curve).}
\label{steadyStates}
\end{figure}

Using the results summarized in Figure \ref{kinkForce} we should find
static solutions of Eq.~(\ref{beam}) near the zeros of $\cos
(2x-0.8159)$. The first six such zeros are at $x=$1.19, 2.76, 4.33, 5.91, 7.48, 9.04. 
We can use Matlab's \emph{fsolve} command on the system $D_{2}^{2}u+V^{\prime }(u)=0$ 
with an initializer that is close to the desired equilibrium
solution in order for \emph{fsolve} to converge to that solution. For the initializer we 
use Eq.~(\ref{steadyInitializer}) again, with $x_0$ close to one of the above zeros.
Note that the value of $x_0$ moves significantly from $x_0=1.19$ in the initializer to 
$x_0=1.30$ in the \emph{fsolve} full solution for the first case (because we are in the
region $x < 1.8$ where the asymptotic 
fit is breaking down); for the other cases, there is negligible change. See Figure \ref{steadyStates} for the
first four equilibrium solutions. Note that there is no further solution with $x_0$ smaller.

\begin{figure}[hbtp]
\begin{center}
\includegraphics[width=0.46\textwidth]{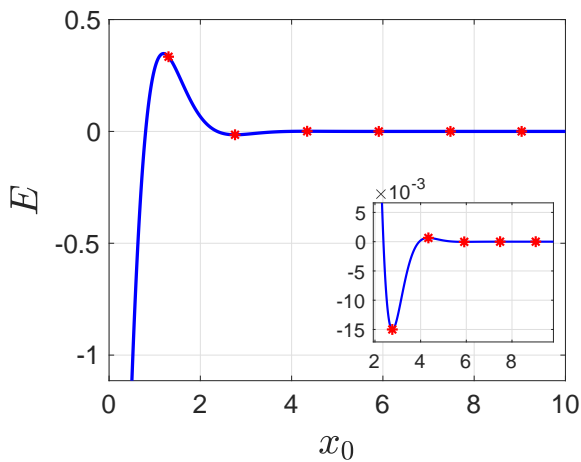}
\end{center}
\caption{Energy vs $x_0$. Blue curve is $-2M(a e^{-2x_0}(\sin(2x_0+b)
+ \cos(2x_0+b)))/4$ with  $M=1.18519$, $a=6.389$ and $b=0.7549$,
which is $\displaystyle 2M\int A_K(x_0)dx_0$ where $A_K(x_0)$ is
defined in Eq.~(\ref{A_k}). The red points are the normalized potential energies of the equilibria 
at $x_0$=1.30, 2.76, 4.33, 5.91,7.48, 9.04, the first four of which are shown in Figure \ref{steadyStates}.}
\label{energy}
\end{figure}

Furthermore, we should be able to predict the local stability of each of the
equilibrium solutions using Equation (\ref{akOde}). When $x_0$ is in one of the intervals 
$(0,1.30)$, $(2.76,4.33)$, $(5.91,7.48)$ the acceleration $A_{AK}(x_0)$ of the antikink is negative 
(since the kink acceleration is positive there), and when $x_0$ is in one of the intervals 
(1.30,2.76), (4.33,5.91), (7.48,9.04), $A_{AK}(x_0)$ is positive. Thus a kink starting at rest in the 
interval (0,1.30) or (1.30,2.76) will tend to start moving away from $x=1.30$ indicating a saddle 
point in the phase portrait of Equation (\ref{akOde}). A kink starting at rest in (1.30,2.76) or 
(2.76,4.33) will tend to move towards $x=2.76$ indicating a center at that point. Similarly we 
expect saddles at $x=4.33$ and $7.48$ and centers at $x=5.91$ and $x=9.04$. 

For another perspective on the equilibrium solutions shown in 
Figure \ref{steadyStates}, their stability, and their relationship to
the acceleration curve in Figure \ref{kinkForce}, we proceed as
follows. Multiplying the acceleration term in Eq.~(\ref{akOde}) by
$2M$ ($M=1.18519$, calculated numerically) and then integrating it
gives the potential energy graph. We can then calculate the field-theoretic
potential energy of each of the equilibrium solutions as $\int_{-\infty
}^{\infty }(\frac{1}{2}u_{xx}^{2}+V(u))dx$ (after which we normalize by subtracting the
limiting value at infinite separation, about $2.099666$, to make the limiting
value of the interaction potential zero). We then plot these
points along with the potential energy graph in Figure
\ref{energy}. We see that the potential energy of the equilibria
occurs at the maximum or minimum points on the potential energy
graph. The leftmost point, while still lying on the graph, is not quite
at the nearby maximum. This is not unexpected, as it was shown in
Figure \ref{kinkForce} that a good fit to the asymptotic data does not begin
until about $x_0=2$. The maxima are the unstable equilibria
(saddles), and the minima are the stable equilibria (centers), as expected.

\begin{figure}[hbtp]
\begin{center}
\includegraphics[width=0.45\textwidth]{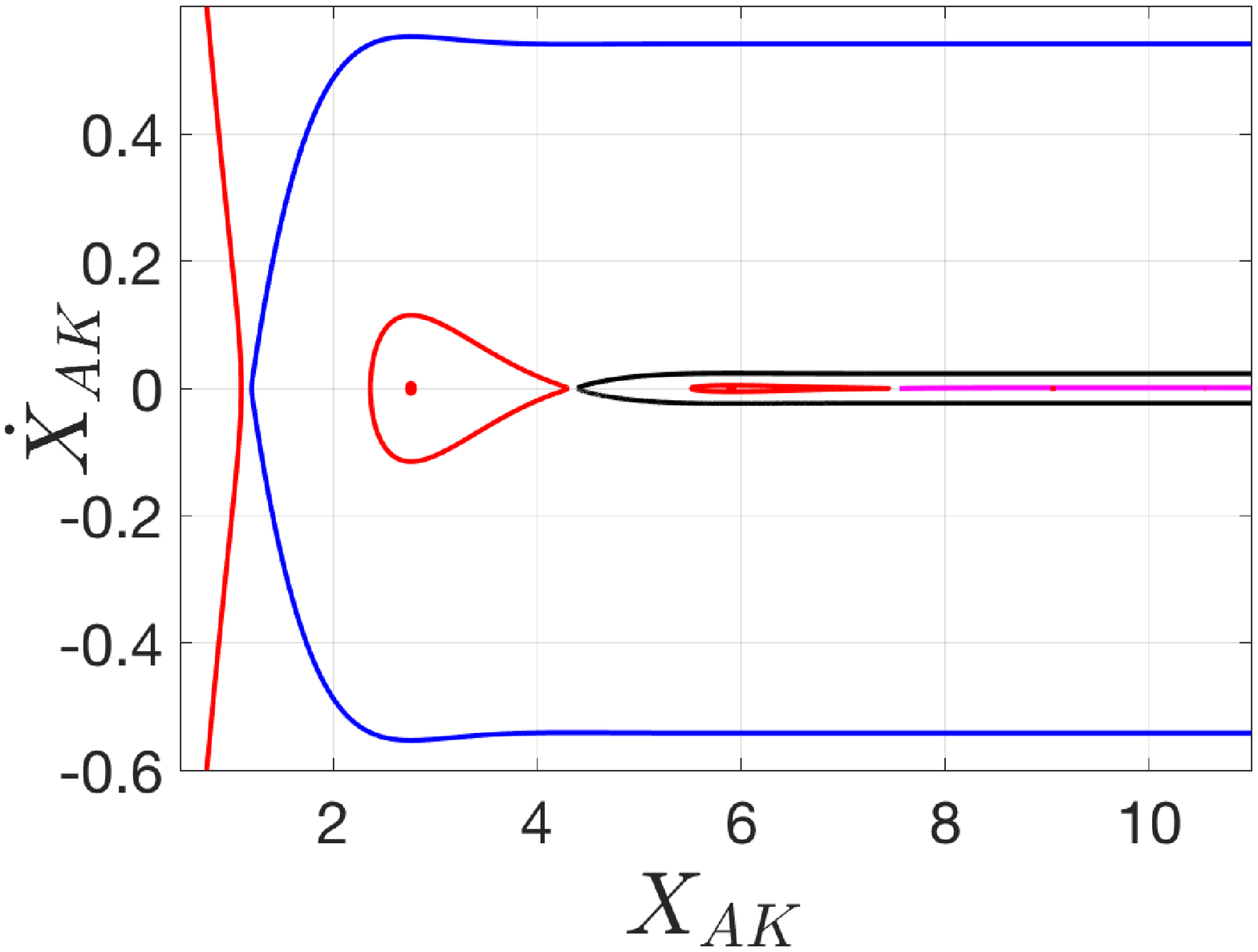}
\includegraphics[width=0.45\textwidth]{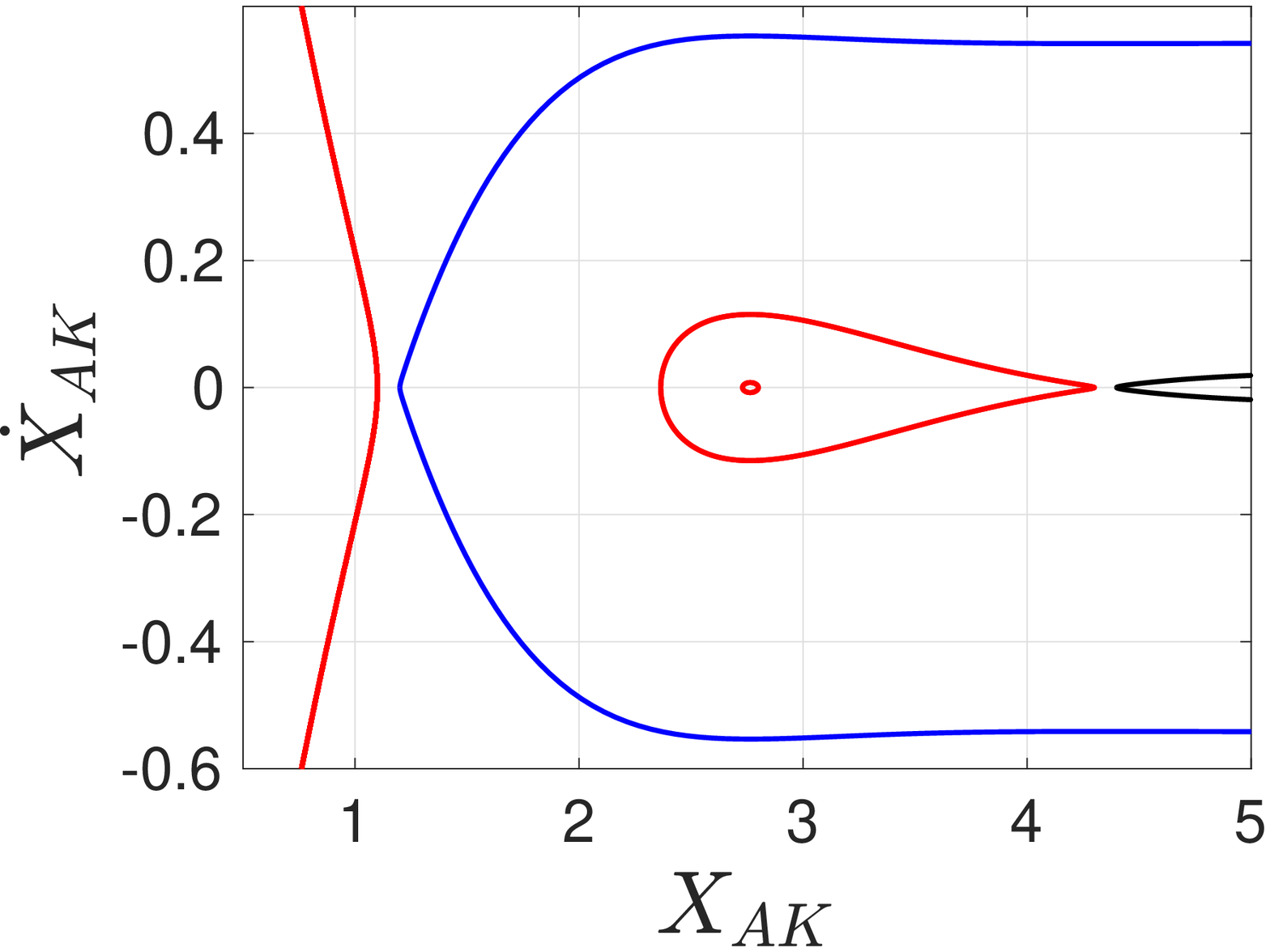} \\
\includegraphics[width=0.45\textwidth]{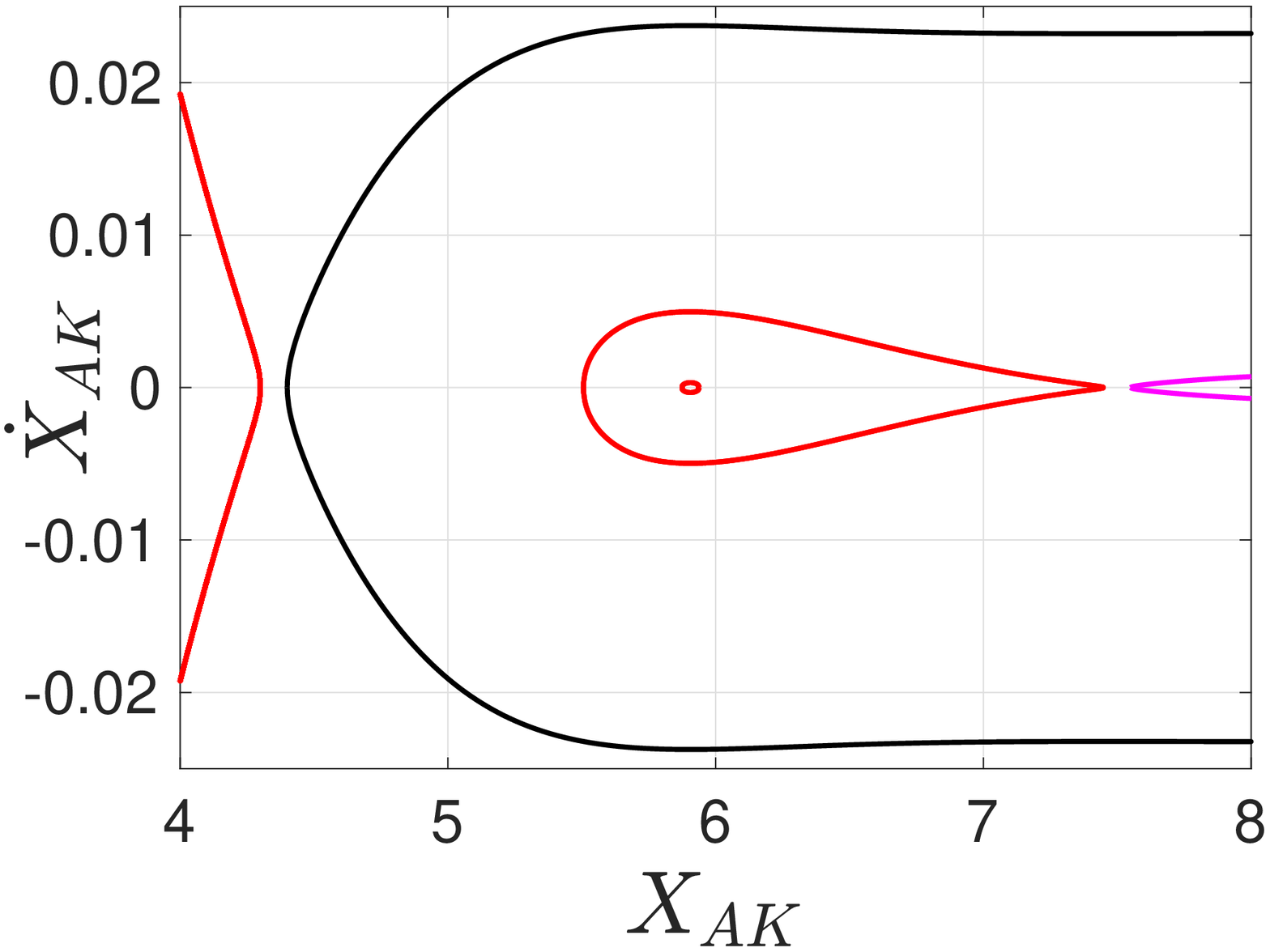} 
\includegraphics[width=0.45\textwidth]{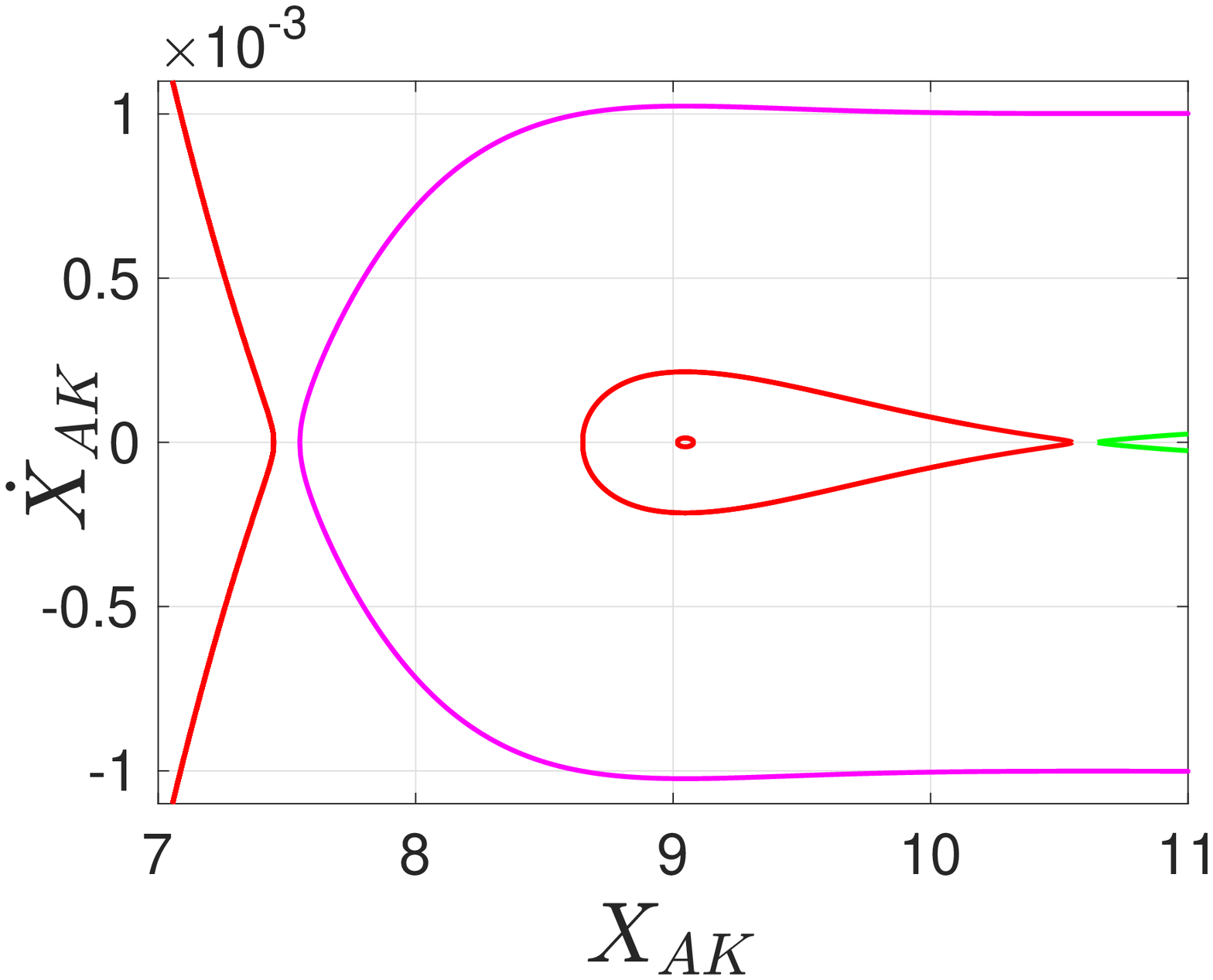}
\end{center}
\caption{Phase portrait of Equation (\ref{akOde}) at different scales;
colors in the top left panel correspond to colors in the other figures. Top left: Overall portrait 
for $0.5\leq X_{AK}(x_0)\leq 11$ and $-0.6\leq \dot{X}_{AK}(x_0)\leq 0.6$. Top right: Zoom to 
$0.5\leq X_{AK}(x_0)\leq 5$ and $-0.6\leq \dot{X}_{AK}(x_0)\leq 0.6$. Bottom left: Zoom to 
$4\leq X_{AK}(x_0)\leq 8$ and $-0.025\leq \dot{X}_{AK}(x_0)\leq 0.025$. Bottom right: Zoom 
to $7\leq X_{AK}(x_0)\leq 11$ and $-0.0011\leq \dot{X}_{AK}(x_0)\leq 0.0011$.}
\label{phasePortrait}
\end{figure}

Figure \ref{phasePortrait} shows the phase portrait of 
Eq.~(\ref{akOde}) for different intervals on the $X_{AK}(x_0)$ axis and
different scales along the $\dot{X}_{AK}(x_0)$ axis, which verifies
the existence of the centers and saddles at the values given above. We
also note the apparently self-similar nature of the phase portrait,
exhibiting a qualitative repetition at progressively smaller scales,
but do not pursue this further here.

\begin{figure}[hbtp]
\begin{center}
\includegraphics[width=0.46\textwidth]{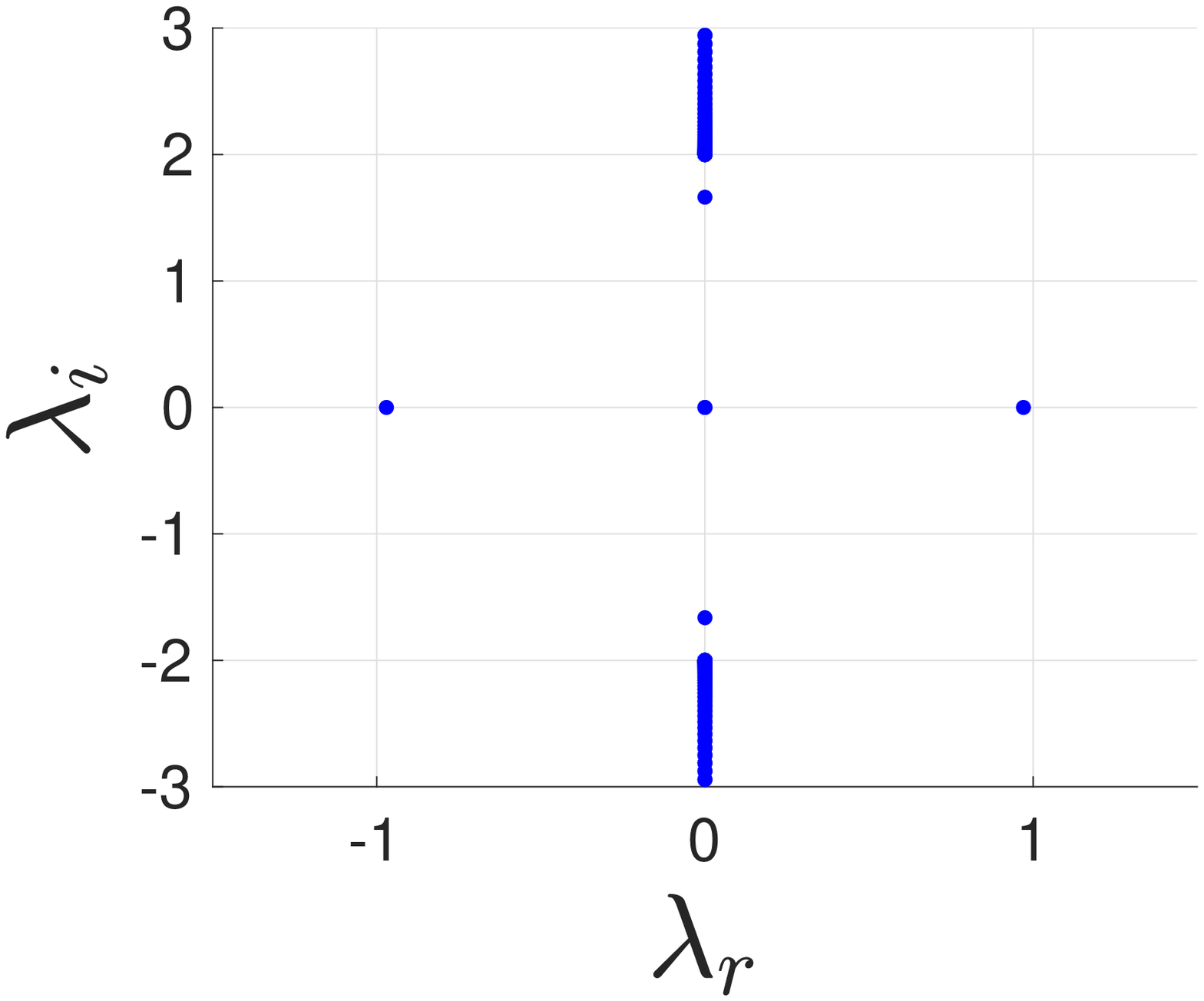}
\includegraphics[width=0.46\textwidth]{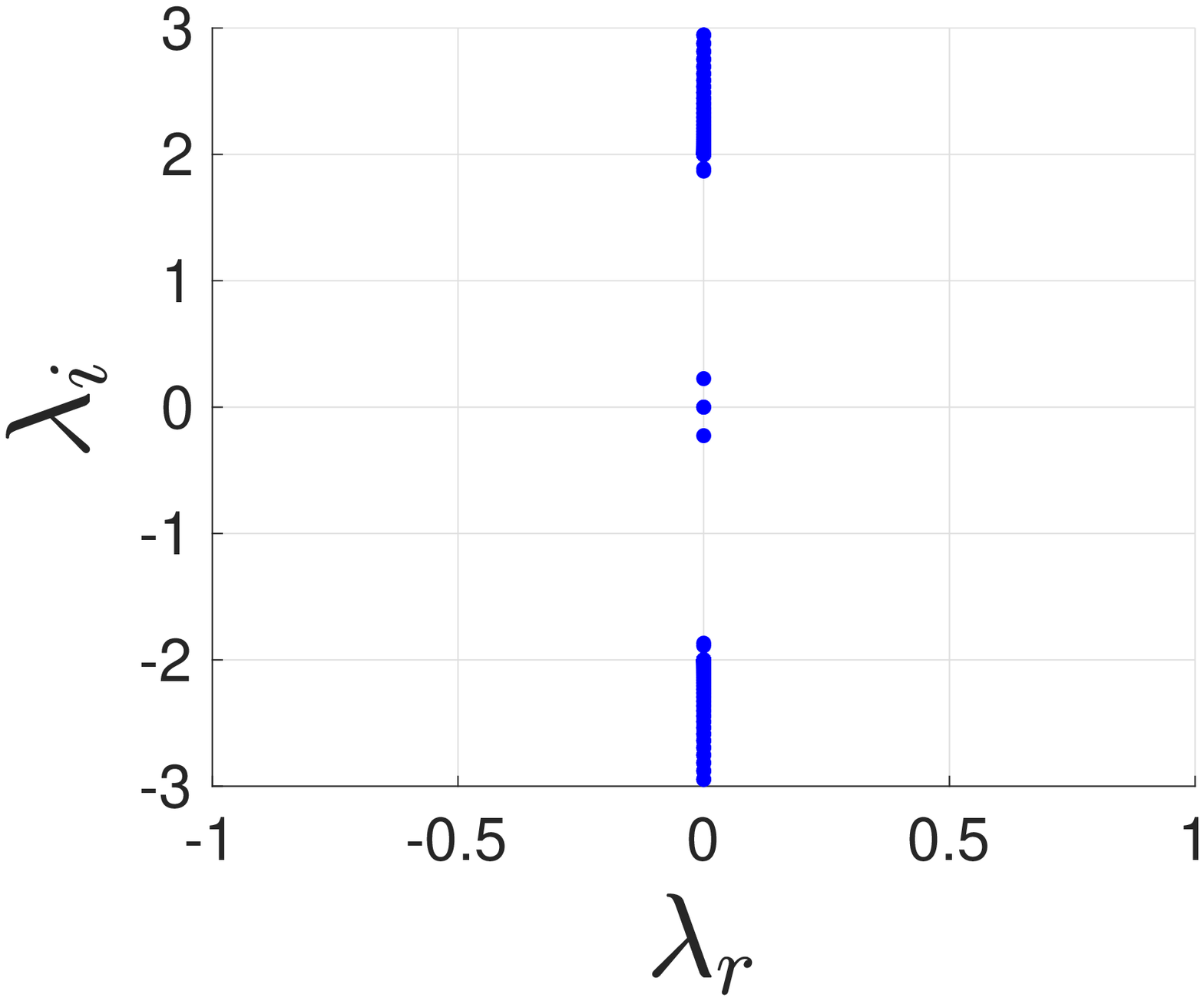}\\
\includegraphics[width=0.46\textwidth]{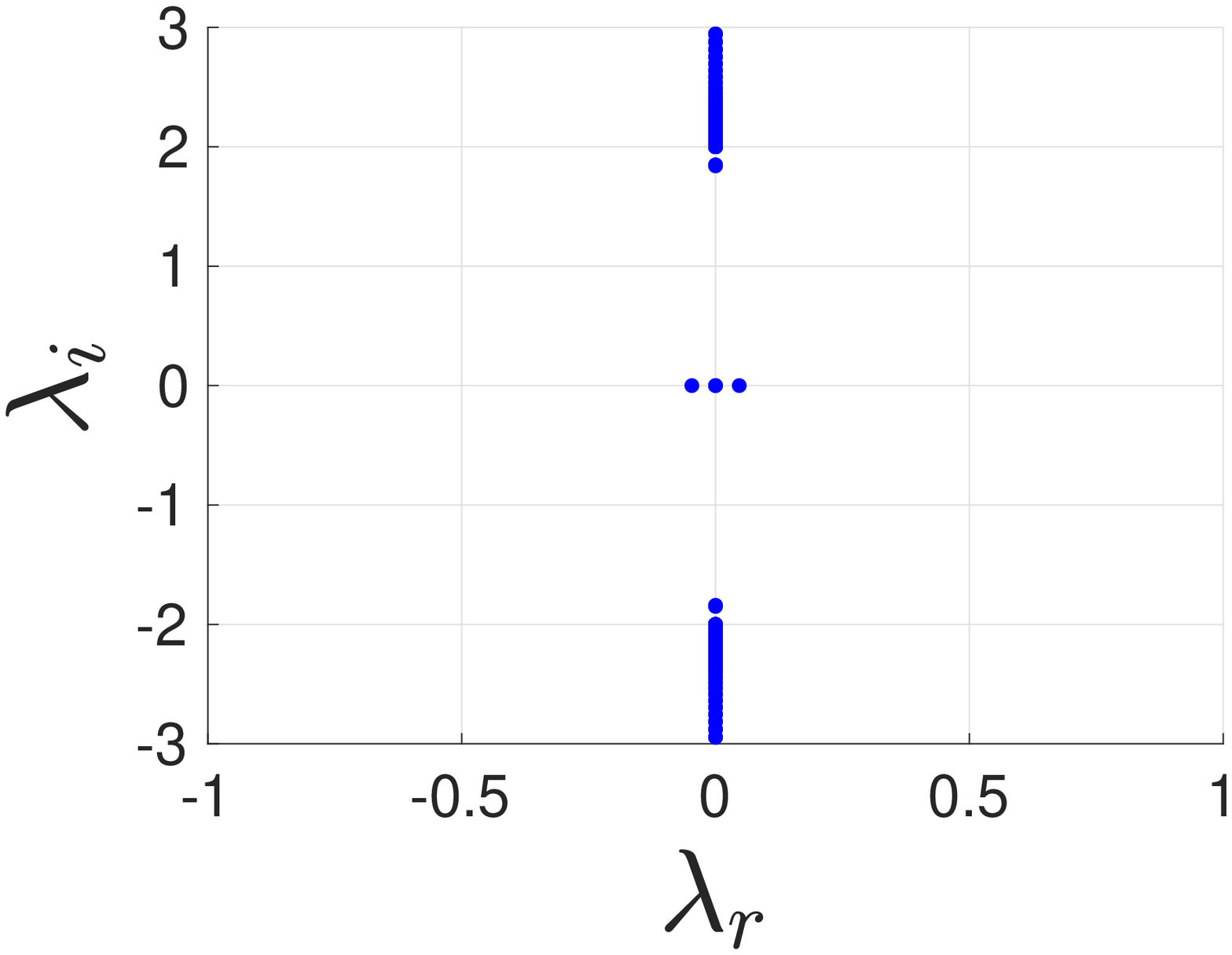}
\includegraphics[width=0.46\textwidth]{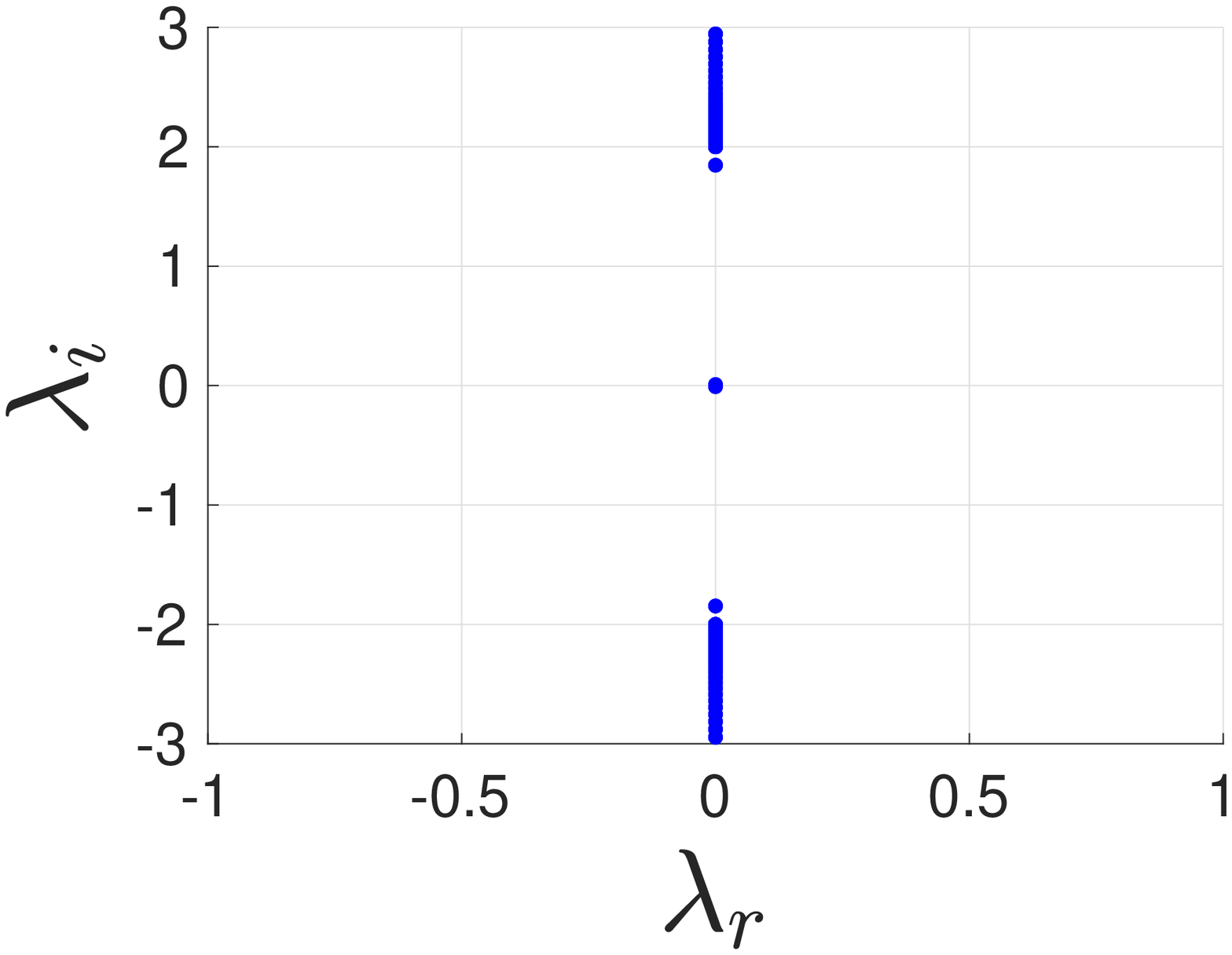}
\end{center}
\caption{ The spectral plane $(\lambda_r, \lambda_i)$ of eigenvalues $\lambda = \lambda_r + i\lambda_i$ 
of oscillations around the equilibria at $x_0=1.30$ (top left), $x_0=2.76$ (top right), $x_0=4.34$ 
(bottom left), $x_0=5.91$ (bottom right).}
\label{eigenvalues}
\end{figure}

For the PDE, Eq.~(\ref{beam}), we expect that for
$x_0=1.30$ and $x_0=4.34$, the equilibrium solutions shown in Figure
\ref{steadyStates} are locally unstable and those for $x_0=2.76$
and $x_0=5.91$ are locally stable. This is confirmed by Figure
\ref{eigenvalues}, where the spectral plots $(\lambda_r,\lambda_i)$
are shown for the eigenvalues $\lambda=\lambda_r+i \lambda_i$ of the 
linearized field equation. Using the expansion
$u(x,t)=u_0(x)+ \epsilon e^{\lambda t} w(x)$ around an equilibrium solution
$u_0(x)$ and solving for the eigenvalues $\lambda$ and eigenvectors $w$, 
we conclude that the equilibrium is stable for $x_0=2.76$ and $x_0=5.91$, as
all eigenvalues are imaginary, and that it is unstable for
$x_0=1.30$ and $x_0=4.34$, as in that case there is one real eigenvalue pair.
The lowest non-zero imaginary eigenvalue in the former case, as well
as the single nonvanishing real pair in the latter case correspond to
the mode associated with the relative motion of the kink
and antikink centres, leading to stable oscillations in the former case and
unstable sliding away in the latter. The vanishing pair of eigenvalues
is associated with the rigid translation of the kink-antikink pair, which
is energy-neutral and whose eigenvector $w = u_x$ is the translation zero mode.
Lastly, we note the presence of another nontrivial imaginary eigenvalue below 
the phonon band of spatially extended modes
which appears to be analogous to the
well-known internal excitation mode of the $\phi^4$ Klein-Gordon 
kink~\cite{Sugiyama,Campbell,Ann}.

\subsection{Kink-Antikink Interactions with Non-Zero Initial Velocities}

In \cite{beam1} a kink and antikink were sent towards each other at
various initial velocities $v_{in}$  and the outgoing velocity
$v_{out}$ was recorded. For $v_{in}$ up to a critical value of
approximately $0.5108$ it was found that the solitons rebound
elastically ($v_{out} =v_{in}$), and for velocities greater than a
second critical value of approximately $0.5896$ the solitons interact
once and then separate with $v_{out}<v_{in}$. Between these two
critical values, the solitons get trapped and form a bion
state. Furthermore, the kink and antikink appeared to approach and
oscillate about a steady state when $v_{in}$ was very close to the
first or second critical value. We can now use the results of the
present paper to explain some of these observations.

Using Eq. (\ref{akOde}) of Section \ref{steadySec} we can make some predictions 
about how the kink and antikink will interact, provided their separation does not get 
close to zero. From Figure \ref{phasePortrait} we predict that for $x_0=10$, the behaviour 
depends on the initial velocity. The initial velocity that creates the blue trajectory is about
$v_{\mathrm{in}}=-0.54$ as can be seen in the first two panels. Similarly, the
initial velocities that create the black (magenta) trajectories are about
$v_{\mathrm{in}}=-0.023$ $(-0.001)$ as can be seen from the third (final) panels. Thus in
all cases, when $0.001<|{v_{\mathrm{in}}}|<0.54$ we expect that the
kink and antikink will approach each other up to a certain point, then
reverse direction under the influence of one of the saddle points,
eventually attaining the velocity
$v_{\mathrm{out}}=|v_{\mathrm{in}}|$, so there is no loss of energy. The
minimum kink-antikink separation depends on which saddle ``turns back''
the trajectory.  Also, there will be a jump in the minimum separation
near each of the $v_{\mathrm{in}}$ values given above.

For $x_0=10$ there is a further possibility. If $|v_{\mathrm{in}}|<0.001$, the trajectory will orbit 
the center at $x=9.04$. Thus in the PDE simulation, we should see the kink and antikink both 
oscillating for all time. In this case there is no $v_{\mathrm{out}}$. Oscillations can occur around 
any center with a smaller $x_0$ value, but this requires the kink and antikink to start closer together.

\begin{figure}[hbtp]
\begin{center}
\includegraphics[width=0.46\textwidth]{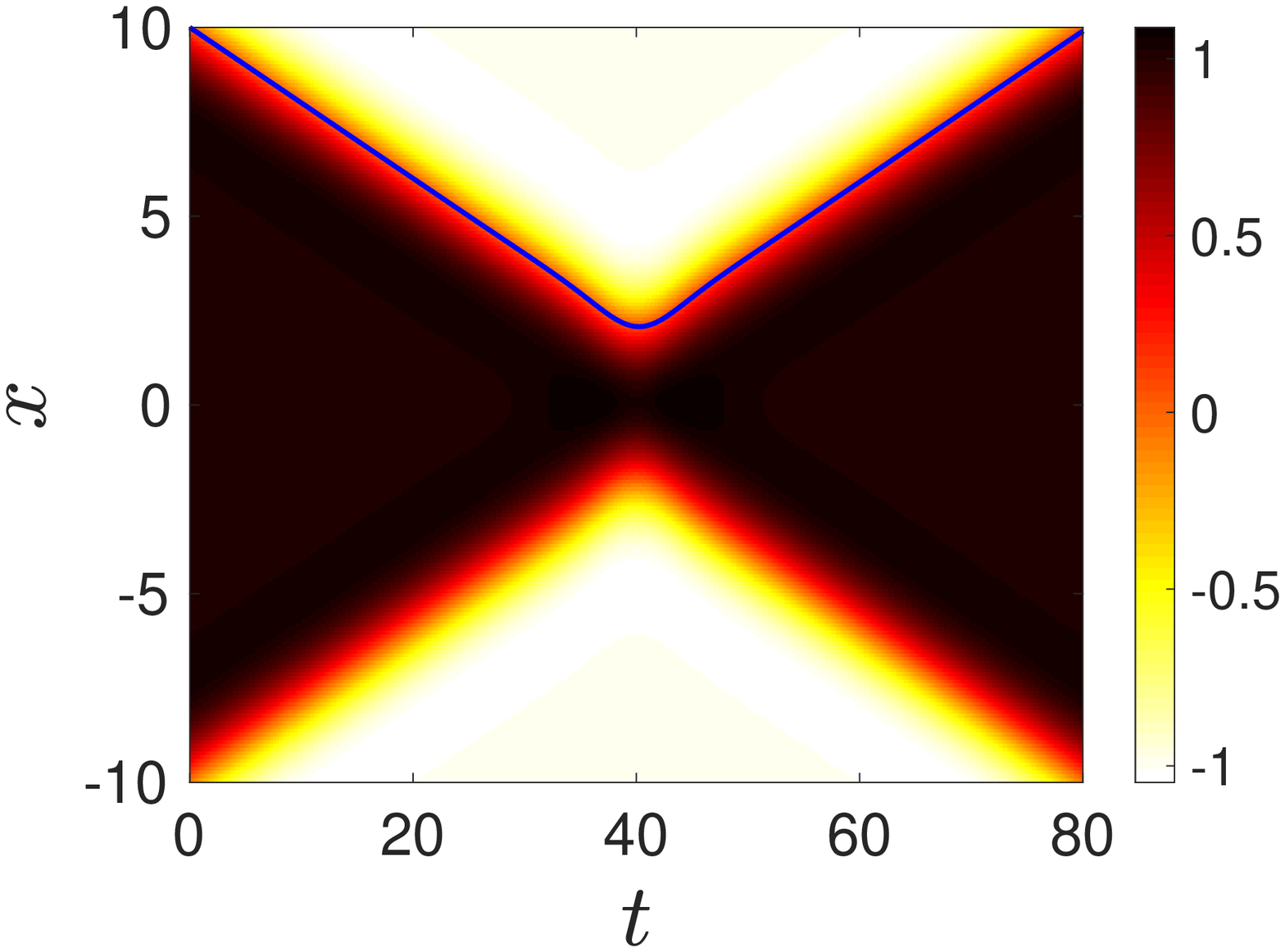}
\includegraphics[width=0.46\textwidth]{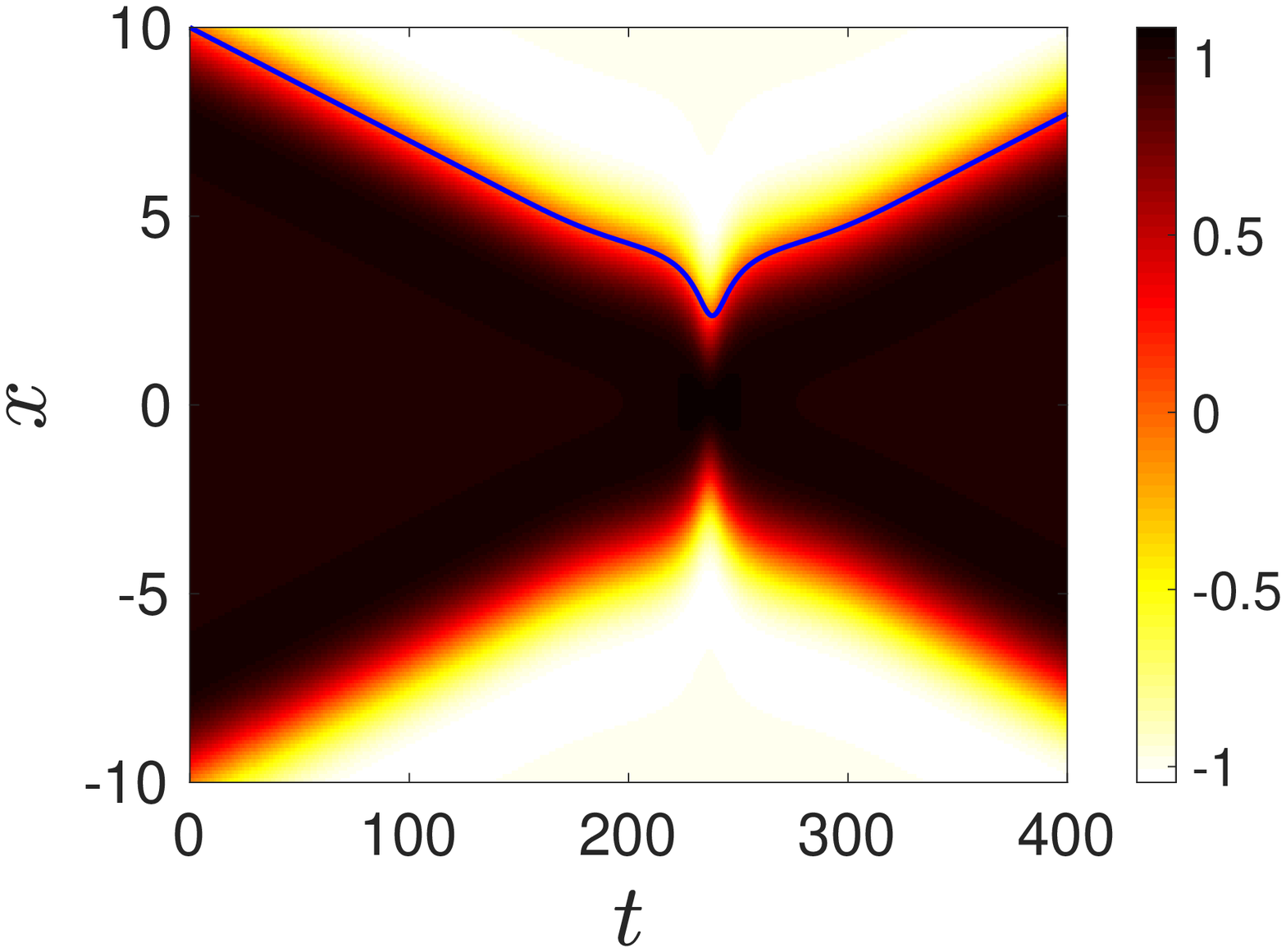}\\
\includegraphics[width=0.46\textwidth]{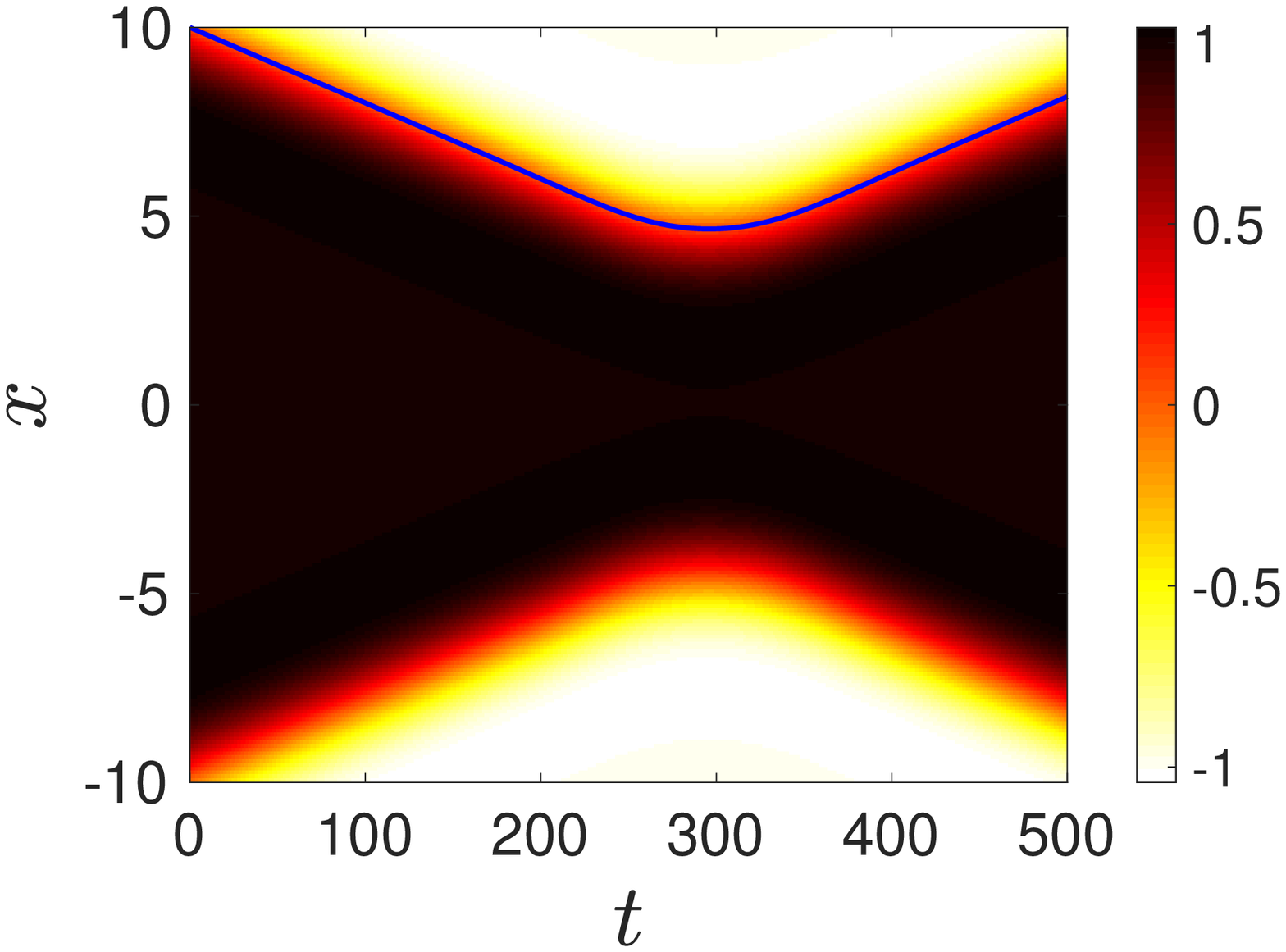}
\includegraphics[width=0.46\textwidth]{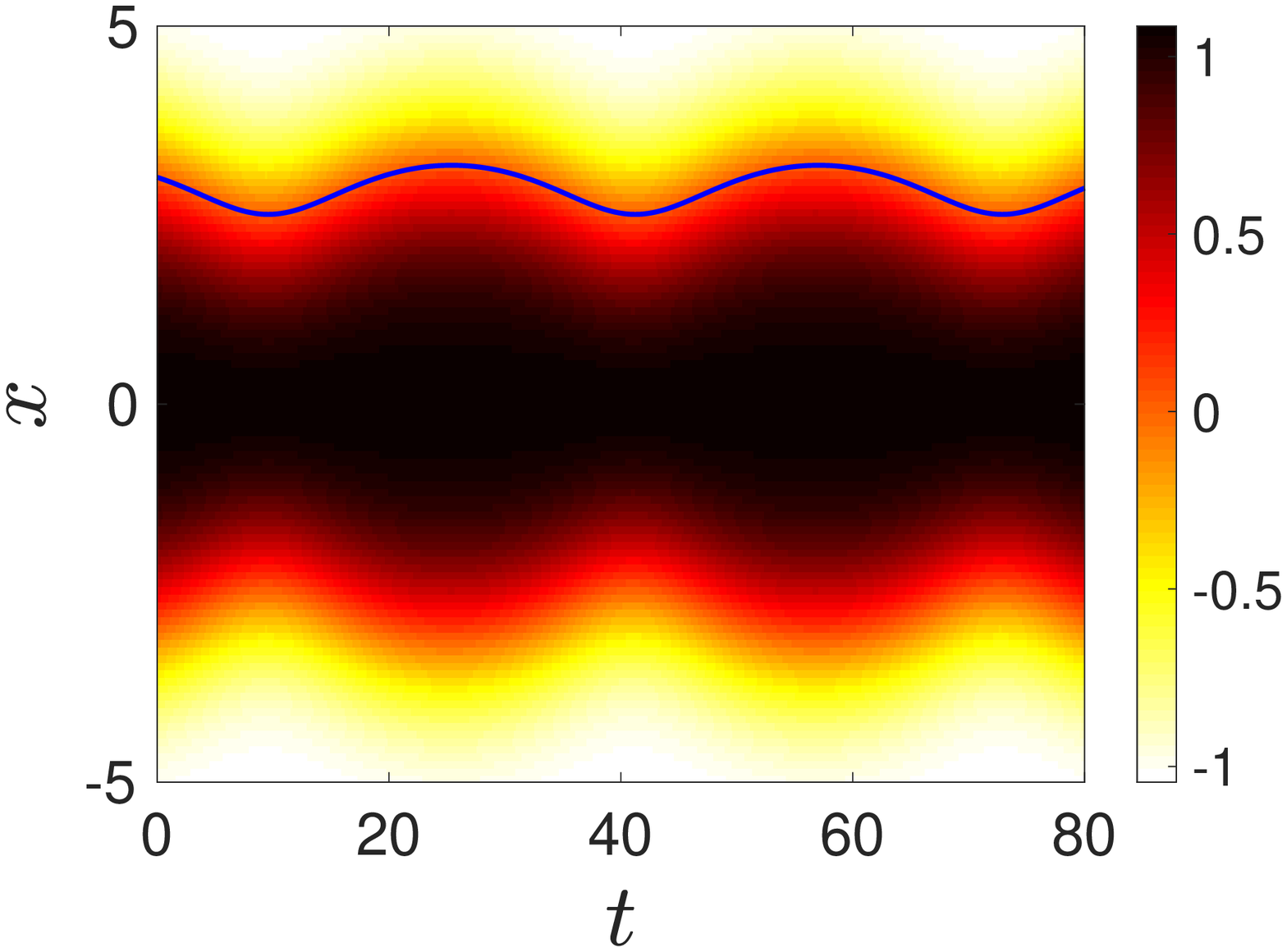}
\end{center}
\caption{Comparisons of the PDE contour plot of the displacement field 
$u(x,t)$ and the ODE trajectory solving Eq. \ref{akOde} (blue solid curve). 
Upper left: $x_0=10$, $|v_{\mathrm{in}}|=0.2$. Upper right: 
$x_0=10$, $|v_{\mathrm{in}}|=0.03$. Lower left: $x_0=10$, $v_{\mathrm{in}}=0.02$. 
Lower right: $x_0=3$, $|v_{\mathrm{in}}|=0.05$.}
\label{contourOde}
\end{figure}

We turn now to the PDE simulations to see if our predictions based on the simple ODE 
model Eq. (\ref{akOde}) hold. Figure (\ref{contourOde}) shows contour plots of the PDE, 
corresponding to three $v_{\mathrm{in}}$ values for a separation half-distance of $x_0=10$, 
and one $v_{\mathrm{in}}$ value with $x_0=3$. For each case, we also plot the solution to 
Eq. (\ref{akOde}) in blue on top of the contour plot. In all cases, the simple model correctly 
predicts the motion of the center of the antikink in the PDE simulation. 
Note that in moving from the upper left panel, to the upper right
panel, to the lower left panel of Figure \ref{contourOde} we see that
the minimum approach distance transitions from about $x=2$ to about $x=5$;
this corresponds to a transition from a phase-plot trajectory which is
inside the blue trajectory in Figure \ref{phasePortrait} to a
trajectory inside the black trajectory in that figure, bypassing the
saddle at $x=4.33$. Further reductions in the value of
$|v_{\mathrm{in}}|$ would show this process repeating, with the
phase-plot trajectory bypassing the saddle at $x=7.48$ (now inside the
magenta trajectory) resulting in a minimum approach distance 
between approximately $7.48$ and $8.5$. If $|v_{\mathrm{in}}|$ is reduced even
further, the result is a trajectory around the center at $x=9.04$. (The time for
the PDE simulation to show oscillations about the center at $x=9.04$ is rather large.) 
The bottom right panel shows a trajectory that encloses a different center,
the one at $x=2.76$.

The left boundary of the interval where multiple bounces occur,
$|v_{\mathrm{in}}|=0.5109$, corresponds to the $v_{\mathrm{in}}$
value that creates the trajectory that approaches the saddle at
$x=1.3$. Note that this is somewhat inconsistent with the phase
portraits in Figure \ref{phasePortrait} (top level) which indicates a
value of about $v_{\mathrm{in}}=-0.54$ (vertical coordinate of the
blue trajectory at $x_0=10$). As noted previously, this is due to the
fact that the saddle in the model given by Eq. (\ref{akOde}) is at
$x=1.19$ but the unstable equilibrium in the PDE model is at $x_0=1.3$
(recall that for $x<1.8$ the ODE model loses accuracy). Nevertheless,
the asymptotic analytical formulation of Eq.~(\ref{prediction}) and
the corresponding numerical finding of Eq.~(\ref{A_k}) provide a
particularly useful energy landscape for kink-antikink collisions in 
our beam model.

\section{Conclusions and Future Challenges}

In the present work, we have examined the kink-antikink interaction in 
a nonlinear beam model with a $\phi^4$ potential, i.e., a cubic nonlinearity. 
We have deployed an asymptotic methodology based on~\cite{Manton_nuclear}
to find the force acting between a kink and antikink, and hence their accelerations. 
The oscillatory tails of these structures (as discovered in~\cite{beam1}) imply 
that there is an exponentially modulated, spatially oscillatory force 
alternating between regions of attraction and repulsion. The saddles 
and centers of the effective dynamics with one degree
of freedom, which are stationary points of an effective potential,
are confirmed through PDE computations. The predicted kink and antikink 
accelerations are also confirmed by direct numerical computations using 
the PDE. In addition, the collision dynamics implied by the kink-antikink potential energy
landscape is found to be in good agreement with direct PDE time evolution simulations, 
except when the separation is very small.

There are numerous directions in which one could extend this work.
We did not yet study the interplay of the translational motion of the
kink and antikink with the internal mode that kinks in this nonlinear 
beam model appear to possess, according to our stability analysis. 
It would also be interesting to relate our work to recent studies 
of experimentally relevant pure-quartic solitons, and of the effect of mixed 
second and fourth derivatives in the NLS realm~\cite{pqs,pqs2}. 
Lastly, it would be interesting to seek models in higher dimensions
where other types of solitary waves, for example, vortices, have a potential 
energy landscape with multiple stationary points. 

\section*{Acknowledgements} NSM is partially supported by STFC
consolidated grant ST/P000681/1.
This material is based upon work supported by the US National Science
Foundation under Grants No. PHY-1602994 and DMS-1809074
(PGK). PGK also acknowledges support from the Leverhulme Trust via a
Visiting Fellowship and thanks the Mathematical Institute of the University
of Oxford for its hospitality during part of this work.

\end{document}